# A Graph Theoretic Approach to Power System Vulnerability Identification

Reetam Sen Biswas, *Student Member, IEEE*, Anamitra Pal, *Senior Member, IEEE*, Trevor Werho, *Member, IEEE*, and Vijay Vittal, *Fellow, IEEE*

***Abstract*— During major power system disturbances, when multiple component outages occur in rapid succession, it becomes crucial to quickly identify the transmission interconnections that have limited power transfer capability. Understanding the impact of an outage on these critical interconnections (called saturated cut-sets) is important for enhancing situational awareness and taking correct actions. This paper proposes a new graph theoretic approach for analyzing whether a contingency will create a saturated cut-set in a meshed power network. A novel feature of the proposed algorithm is that it lowers the solution time significantly making the approach viable for real-time operations. It also indicates the minimum amount by which the power transfer through the critical interconnections should be reduced so that post-contingency saturation does not occur. Robustness of the proposed algorithm for enhanced situational awareness is demonstrated using the IEEE-118 bus system as well as a 17,000+ bus model of the Western Interconnection (WI). Comparisons made with different approaches for power system vulnerability assessment prove the utility of the proposed scheme for aiding power system operations during extreme exigencies.**

## NOMENCLATURE

| | |
|---|---|
| $c_l^{FT}$ | Directed weight associated with edge $e_l$ from vertex $v_l^F$ towards $v_l^T$ in the latent capacity graph ($\mathcal{C}$). |
| $c_l^{TF}$ | Directed weight associated with edge $e_l$ from vertex $v_l^T$ towards $v_l^F$ in the latent capacity graph ($\mathcal{C}$). |
| $\boldsymbol{C}_1$ | The set of vertices contained in cluster 1. |
| $\boldsymbol{C}_2$ | The set of vertices contained in cluster 2. |
| $C_P$ | Maximum extra flow that can be transferred along path $\boldsymbol{P}$ from a source vertex towards a sink vertex. |
| $D_l$ | Active power withdrawn at a sink vertex $v_l \in \boldsymbol{L}$. |
| $e_l$ | $l^{th}$ edge in the edge set $\boldsymbol{E}$. |
| $\boldsymbol{E}$ | A set containing all edges of the power network. |
| $f_l$ | A directed weight associated with edge $e_l$ from vertex $v_l^F$ towards vertex $v_l^T$ in the flow graph ($\mathcal{F}$). |
| $f_l^A$ | Flow in edge $e_l$ for the network flow solution $A$. |
| $F_P$ | The flow injected along path $\boldsymbol{P}$. |
| $\boldsymbol{G}$ | A set containing the locations of generator buses. |
| $I_g$ | Active power injected at a source vertex $v_g \in \boldsymbol{G}$. |
| $k$ | Total number of edges in cut-set $\boldsymbol{K}$. |
| $\boldsymbol{K}$ | Any cut-set in the power network. |
| $\boldsymbol{K}_i$ | $i^{th}$ cut-set associated with edge $e_l \in \boldsymbol{E}$. |
| $\boldsymbol{K}_{crit}$ | Limiting critical cut-set for edge $e_l \in \boldsymbol{E}$. |
| $\boldsymbol{L}$ | A set containing all the load buses. |
| $n$ | Total number of indirect paths for edge $e_l$. |
| $\boldsymbol{P}$ | This is a path (sequence of edges) from a source vertex to a sink vertex in the graph $\mathcal{G}$. |
| $P_G^1$ | Total active power generation in cluster $\boldsymbol{C}_1$. |
| $P_G^2$ | Total active power generation in cluster $\boldsymbol{C}_2$. |
| $P_L^1$ | Total active power demand in cluster $\boldsymbol{C}_1$. |
| $P_L^2$ | Total active power demand in cluster $\boldsymbol{C}_2$. |
| $\Delta P^1$ | Net active power injection in cluster $\boldsymbol{C}_1$. |
| $\Delta P^2$ | Net active power injection in cluster $\boldsymbol{C}_2$. |
| $P_K$ | Total active power to be transferred across cut-set $\boldsymbol{K}$. |
| $r_l$ | Rating of edge $e_l \in \boldsymbol{E}$. |
| $R_K$ | Total active power transfer capacity of cut-set $\boldsymbol{K}$, excluding edge $e_l$, examined by feasibility test (FT). |
| $T_l^i$ | Transfer margin of the $i^{th}$ saturated cut-set, associated with edge $e_l$. |
| $T_l$ | Transfer margin of the limiting critical cut-set associated with edge $e_l$. |
| $TC_l$ | Total additional active power transfer capability of the indirect paths of edge $e_l$. |
| $v_g$ | A vertex that has a source (or generator). |
| $v_l$ | A vertex that has a sink (or load). |
| $v_l^F$ | The "from vertex" of edge $e_l$. |
| $v_l^T$ | The "to vertex" of edge $e_l$. |
| $\boldsymbol{V}$ | A vertex set containing all buses of the power network. |
| $x$ | Total number of cut-sets associated with $e_l$. |
| $y$ | Total number of saturated cut-sets associated with $e_l$. |
| $z$ | A variable denoting impedance of a branch. |
| $\mathcal{G}$ | An undirected weighted graph of the power network. |
| $\mathcal{F}$ | A directional flow graph of the power network. |
| $\mathcal{C}$ | A bidirectional latent capacity graph of the network. |

This work was supported in part by the Power System Engineering Research Center (PSERC) Grants S-74 and S-87.

The authors are associated with the School of Electrical, Computer, and Energy Engineering, Arizona State University (ASU), Tempe, AZ 85287, USA. (E-mail: rsenbisw@asu.edu, anamitra.pal@asu.edu, twerho@asu.edu, vijay.vittal@asu.edu).

## I. INTRODUCTION

ANALYSIS of major blackouts has indicated that they involve successive outages of power system assets [1]. For example, the 1977 New York City blackout was caused by the loss of 11 transmission lines in 52 minutes. The Federal Electricity Regulatory Commission (FERC) reported that one of the causes of the blackout was "the failure to recognize that a critical interconnection to the west was effectively unavailable" [2]. More recently, the initiating event for the 2011 U.S. Southwest blackout was the loss of the 500 kV Hassayampa-North Gila (H-NG) line, which then triggered a sequence of events that resulted in the blackout of San Diego [3]. Werho et al. stated that a critical interconnection does not necessarily refer to a single line whose status can be monitored [4]; i.e., *a critical interconnection can consist of multiple lines*. Therefore, real-time vulnerability assessment for enhanced situational awareness of a power system that is suffering from multiple outages is a challenging task [5], [6].

The traditional approaches for improving situational awareness are based on steady-state contingency analyses techniques that solve AC or DC power flows [7]-[11]. These techniques cannot detect transient/dynamic stability related violations but can identify branch overloads and voltage violations. However, power flow-based contingency analysis (CA) is not fast enough to perform an *exhaustive N*-1 real-time contingency analysis (RTCA) [7]. Therefore, power utilities select a *subset* of the contingencies for evaluation based on some pre-defined criteria [9], [10]. In [11], Huang et al. stated that the size of this subset has considerable impact on RTCA solution: a *large subset is computationally burdensome, while a small subset might miss critical scenarios*. This can be a problem for real-time operations during extreme exigencies when multiple outages occur in rapid succession [4].

For managing extreme event conditions, a variety of approaches that can identify vulnerabilities quickly have been proposed; these include statistical analyses ([12]-[16]), graph theoretic analyses ([4], [17]-[27]), and linear sensitivity-based analyses ([28]-[33]). These types of analyses are suitable for exhaustive *N*-1 and potentially *N-X* evaluations. The proposed graph theoretic approach also belongs to this category of analyses as it enhances situational awareness for real-time operations. A brief overview of these other techniques ([12]-[33]) that belong to this category is provided below.

Dobson et al. in [12], [13] obtained statistics of cascading line outages from utilities to understand how cascades initiate and propagate in the power system. In [14], Rezaei et al. estimated the risk of cascading failure with an algorithm called *random chemistry*. In [15], Rahnamay-Naeini et al. performed probabilistic analysis to understand the dynamics of cascading failures. In [16], Hines et al. proposed an *influence graph* model to capture patterns of cascading failures in power systems and validated the model using historical data. Instead of relying on prior historical data, which may or may not be relevant for the present scenario, the proposed approach exploits knowledge of the current network conditions to identify the system's critical interconnections, the loss of which might trigger a cascade.

Graph theoretic approaches have found applications in a variety of fields [34]-[36]. Ishizaki et al. summarized the applications of graph theory for power systems modeling, dynamics, coherency, and control [17]. With regards to vulnerability assessment, graph theoretic approaches have focused on the topology and structure of the power system [18]-[27]. In [18], Albert et al. studied the structural vulnerability of the North American power grid using a metric called the *node degree*, which refers to the number of lines connected to a bus. Use of *betweenness indices*, which refer to the number of shortest paths traversing a given element, were explored in [19], [20]. Such purely topological indices do not consider the electrical properties of the power network.

*Modified centrality indices* were used in [21] and [22] to assess the risk of blackouts/brownouts and systemic vulnerabilities, respectively. In [23] and [24], different statistical measures such as the *betweenness indices*, *node-degree*, and *geodesic distance* were used as possible alternatives to power flow techniques to quantify power system vulnerability during *N*-1 contingencies and cascading failures. In [25], Zhu et al. proposed a metric called *risk graph* to better capture the cascade failure vulnerability of the power system. Recently, Beyza et al. in [26] investigated the structural vulnerability of the power system when successive *N*-1 contingencies progressively alter the network structure. These *global vulnerability metrics* (node degree, betweenness indices, modified centrality indices, geodesic distance) describe the vulnerability of the system by a single number. However, such indices do not convey meaningful actionable information to an operator who is trying to prevent the system from collapsing! This is because these metrics do not consider the *physical manifestation* of a vulnerability – a key issue that the proposed research seeks to address.

In [4], Werho et al. used a graph theory-based *network flow algorithm* to identify the cut-set of minimum size between a source-sink pair. A cut-set denotes the set of edges which when removed separates the graph into two disjoint islands; the size of the cut-set refers to the number of edges present in it. If the number of edges contained in the minimum sized cut-set progressively decreases, it indicates a structural weakness between the selected source-sink pair. In [27], Beiranvand et al. presented a novel topological sorting algorithm to screen out *coherent cut-sets*. Coherent cut-sets denote the set of edges that partition the network, such that the power flows in the same direction through all the edges. However, coherent cut-sets may not be the only bottlenecks in a power system, as there may be a cut-set in which the power flows are not unidirectional, but a single outage limits the power transfer through it.

Bompard et al. used power transfer distribution factors (PTDFs) and transmission line capacities for screening out critical contingencies [28], [29]. Line outage distribution factors (LODFs) have been used for quickly detecting an island formation due to a multiple element contingency [30]. Werho et al. used DC power flow based linear sensitivity analysis to detect an island formation due to a contingency [31]. In [32], [33], contingency screening was done using LODFs. These sensitivity indices capture the topological as well as the electrical properties of the power system and are useful for comparison with the proposed approach (e.g., see Table V).

The goal of this paper is to investigate if *cut-sets will become saturated* (i.e., cannot transfer the required amount of power) due to a would-be outage, irrespective of the direction in which power flows through different edges of the cut-set. Such cut-sets (termed saturated cut-sets henceforth) are the system's critical interconnections as they have limited power transfer capability. Essentially, this paper attempts to answer the following question: *How to quickly make operators aware if a new contingency will create saturated cut-sets in a meshed power network, after multiple component failures have occurred in rapid succession*?

## II. Theoretical Background

### A. Graph theoretic terminologies

In graph theoretic terminology, a power system can be represented by an undirected graph $\mathcal{G}(\boldsymbol{V},\boldsymbol{E})$, such that the buses are contained in the vertex set $\boldsymbol{V}$ and the transmission lines and transformers are contained in the edge set $\boldsymbol{E}$. The generators and loads are the sources and sinks, respectively. The set $\boldsymbol{G}$ consists of all vertices where a source is present and the set $\boldsymbol{L}$ consists of all vertices where a sink is present. The power injected at a source $v_g \in \boldsymbol{G}$ is denoted by $I_g$ and the power demand at a sink $v_l \in \boldsymbol{L}$ is denoted by $D_l$. Now, every transmission asset (line or transformer) has an associated capacity called the asset rating. To account for the asset ratings in the undirected graph $\mathcal{G}(\boldsymbol{V},\boldsymbol{E})$, every edge $e_l \in \boldsymbol{E}$ is associated with a weight $r_l$, where $r_l$ denotes the maximum power that can be transferred through edge $e_l$. From the original graph $\mathcal{G}(\boldsymbol{V},\boldsymbol{E})$, we now create two graphs: the *flow graph*, $\mathcal{F}(\boldsymbol{V},\boldsymbol{E})$, and the *latent capacity graph*, $\mathcal{C}(\boldsymbol{V},\boldsymbol{E})$. The flow graph, $\mathcal{F}(\boldsymbol{V},\boldsymbol{E})$, contains information about the power flow through different edges of the network. If $f_l$ units of power flows through edge $e_l$ from vertex $v_l^F$ towards vertex $v_l^T$, a directed weight of $f_l$ is assigned to edge $e_l$ in a direction from $v_l^F$ to $v_l^T$. On the other hand, for edge, $e_l$, the latent capacity graph, $\mathcal{C}(\boldsymbol{V},\boldsymbol{E})$, provides information regarding the extra flow that could be transferred from $v_l^F$ to $v_l^T$, and vice-versa. The weights associated with the edges of $\mathcal{C}(\boldsymbol{V},\boldsymbol{E})$ that provide information regarding the bidirectional latent capacities are given by,

$$\left.\begin{aligned} c_l^{FT} &= r_l - f_l \\ c_l^{TF} &= r_l + f_l \end{aligned}\right\} \tag{1}$$

where, $c_l^{FT}$ is the latent capacity in the direction from $v_l^F$ to $v_l^T$, and $c_l^{TF}$ is the latent capacity in the direction from $v_l^T$ to $v_l^F$.

### B. *Research scope*

Let an edge $e_l$ (transmission line or transformer) connect vertices (buses) $v_l^F$ and $v_l^T$ as shown in Figure 1. Since edge $e_l$ is a single element that joins vertices $v_l^F$ and $v_l^T$, it is called the direct path from vertex (bus) $v_l^F$ to vertex (bus) $v_l^T$. There could be many other electrical paths to transfer power from $v_l^F$ to $v_l^T$. Any path that contains multiple edges (transmission lines or transformers) from $v_l^F$ to $v_l^T$ is an indirect path. Let there be $n$ indirect paths between vertices (buses) $v_l^F$ and $v_l^T$. If all the $n$ indirect paths combined do not have the capacity to reroute $f_l$ units of power that was flowing through the direct path, it implies that the loss of edge $e_l$ would inevitably result in post-contingency overloads. Based on this inference, a graph theory-based network analysis tool is developed in this paper to quickly detect violations of the type where the set of indirect paths do not have the extra capacity to carry the power that was originally flowing through the direct path.

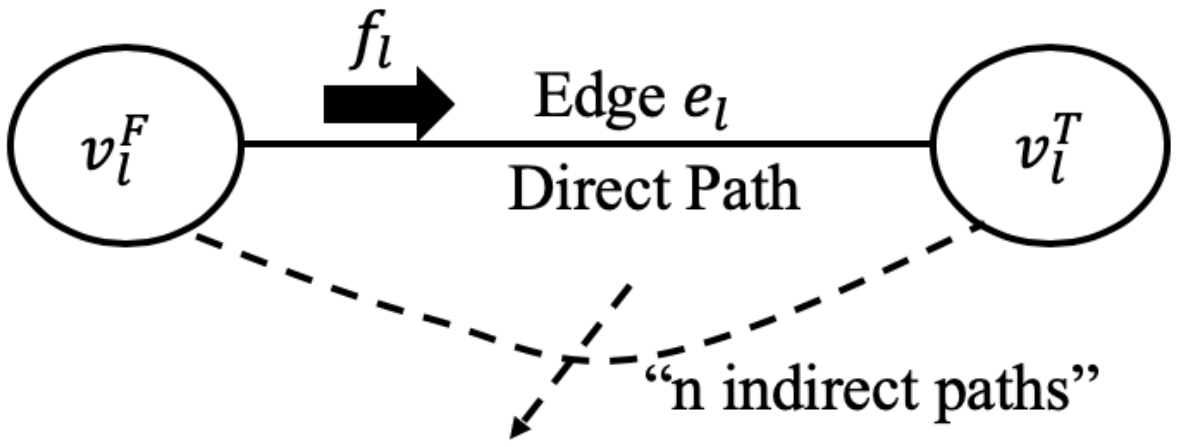

Figure 1: Network connectivity between two vertices (buses)

Contrary to traditional CA studies that detect if an outage causes an overload on the remaining assets of the system [37], the primary goal of this research is to quickly detect if an outage overloads any cut-set of the power system. An overloaded or saturated cut-set is one which transfers power beyond its maximum power transfer capability. Let edges $e_1$, $e_2$,…, $e_k$ belong to cut-set $\boldsymbol{K}$. If the power flowing through the different edges of cut-set $\boldsymbol{K}$ are $f_1$, $f_2$,…, $f_k$, and the ratings of those edges are $r_1$, $r_2$,… $r_k$, then cut-set $\boldsymbol{K}$ will be called a saturated cut-set if the following equation holds true:

$$\sum_{l=1}^{k} f_l > \sum_{l=1}^{k} r_l \text{ , } \forall e_l \in \boldsymbol{K} \tag{2}$$

where, $\sum_{l=1}^{k} f_l$ is the actual power flowing through cut-set $\boldsymbol{K}$ and $\sum_{l=1}^{k} r_l$ is the maximum power that can flow through cut-set $\boldsymbol{K}$ (limited by the ratings of the edges).

At a given time, let us assume that $P_K$ units of power must be transferred through a cut-set, $\boldsymbol{K}$, of a power network. Upon the loss of edge, $e_l$, that belongs to cut-set $\boldsymbol{K}$, if the total power transfer capability of the remaining edges of cut-set $\boldsymbol{K}$ is $R_K$ such that $R_K < P_K$, it implies that the loss of $e_l$ saturates cut-set $\boldsymbol{K}$. In such a situation, edge $e_l$ is termed a *special asset* and the cut-set $\boldsymbol{K}$ is said to be saturated by a negative transfer margin of $R_K - P_K$ due to the loss of the special asset, $e_l$.

Now, let edge $e_l$ be associated with $x$ cut-sets of the network, of which $y$ cut-sets ($y \le x$) become saturated by a negative transfer margin when $e_l$ is lost (implying that $y$ cut-sets of the network are saturated). As the $y$ cut-sets may be saturated by different negative transfer margins, $T_l^i$, $1 \le i \le y$, the objective here is to identify the cut-set that becomes saturated by the numerically largest negative transfer margin (i.e., $T_l = \max\left(\left|T_l^i\right|\right);\ 1 \le i \le y$); this cut-set is henceforth referred to as the *limiting critical cut-set*, $\boldsymbol{K}_{crit}$. Quickly identifying the limiting critical cut-set is important because if appropriate preventive control actions are taken so that the limiting critical cut-set is no longer saturated, the proposed approach, which is very fast, can be repeated multiple times until no limiting critical cut-sets are identified. Note that this paper identifies the limiting critical cut-sets based on the thermal ratings of the different assets and the active power flowing through them (power factor is set to unity for the studies done here). However, the proposed network analysis tool is generic enough to incorporate line ratings obtained from other analyses as well (such as, proxy limits based on power system stability criteria).

## III. Graph Theory Based Network Flow

The graph theory-based network flow algorithm is based on the following assumptions: (1) power injections are known, and (2) losses are negligibly small. Subject to these assumptions, the goal is to generate network flows that can help detect if a contingency saturates a cut-set. The graph theoretic network flow algorithm is based on the principle: *utilize the available generation of the sources (generators) to satisfy the total demand of the sinks (loads), without violating the asset ratings*. The network flows are obtained using **Algorithm I** described below. At the start of the algorithm, edges in $\mathcal{F}(\boldsymbol{V},\boldsymbol{E})$ do not have any weight, while the bidirectional weights of edges in $\mathcal{C}(\boldsymbol{V},\boldsymbol{E})$ are equal to the corresponding asset ratings.

The graph theory-based network flow algorithm obeys the law of conservation of energy, but it relaxes Kirchhoff's voltage law as it does not use impedances *directly* while building the network flows; the impedances are accounted for indirectly through the asset ratings. The flow solution is also non-unique because depending on the order in which the sources and sinks are selected, there could be multiple valid flow solutions. However, since the network boundary conditions do not change (i.e., instantaneous power injections are constant), the power transfer across any cut-set of the network is the same for all valid graph-based flow solutions.

**Algorithm I**: Graph theory-based network flow algorithm

i. Randomly select a source vertex $v_g \in \boldsymbol{G}$ and a sink vertex $v_l \in \boldsymbol{L}$.

ii. Search $\mathcal{C}(\boldsymbol{V}, \boldsymbol{E})$ to traverse the shortest unsaturated path $\boldsymbol{P}$ from $v_g$ to $v_l$ using breadth first search (BFS) [37].

iii. Use $\mathcal{C}$ to find the maximum extra flow, $C_P$, that could be transferred from $v_g$ to $v_l$ through path $\boldsymbol{P}$.

iv. Obtain the flow $F_P$ to be injected in $\mathcal{F}(\boldsymbol{V}, \boldsymbol{E})$ along path $\boldsymbol{P}$ from $v_g$ to $v_l$ as $F_P = \min(I_g, D_l, C_P)$.

v. Update weights of edges in graph $\mathcal{F}$ as $f_l = f_l + F_P$, and in graph $\mathcal{C}$ as per (1), for all the edges that belong to path $\boldsymbol{P}$.

vi. Update available generation and unsatisfied demand at $v_g$ and $v_l$ as $I_g := I_g - F_P$ and $D_l := D_l - F_P$.

vii. Depending upon the values of $I_g$ and $D_l$, update the source and sink vertices in accordance with the following logic:

a. if $I_g \neq 0$ & $D_l \neq 0$, the source and sink vertices are not changed;

b. if $I_g = 0$ & $D_l \neq 0$, a new source vertex, $v_g$, is selected from $\boldsymbol{G}$, keeping the sink vertex, $v_l$, unchanged;

c. if $I_g \neq 0$ & $D_l = 0$, a new sink vertex, $v_l$, is selected from $\boldsymbol{L}$, keeping the source vertex, $v_g$, unchanged.

viii. Repeat Steps (ii) through (vii) until the total power generation satisfies the total power demand.

Let the network graph $\mathcal{G}(\boldsymbol{V}, \boldsymbol{E})$ be split into two clusters $\boldsymbol{C}_1$ and $\boldsymbol{C}_2$ such that $\boldsymbol{C}_1 \cup \boldsymbol{C}_2 = \boldsymbol{V}$ and $\boldsymbol{C}_1 \cap \boldsymbol{C}_2 = \emptyset$. If $P_G^1(P_G^2)$ and $P_L^1(P_L^2)$ be the total generation and total demand in $\boldsymbol{C}_1(\boldsymbol{C}_2)$, then the net generation in $\boldsymbol{C}_1$ is given by $\Delta P_1 = P_G^1 - P_L^1$, while the net generation in $\boldsymbol{C}_2$ is given by $\Delta P_2 = P_G^2 - P_L^2$. Now, cut-set $\boldsymbol{K}$ between clusters $\boldsymbol{C}_1$ and $\boldsymbol{C}_2$ would include only those edges whose one end belongs to $\boldsymbol{C}_1$ and the other end belongs to $\boldsymbol{C}_2$; let the number of edges in cut-set $\boldsymbol{K}$ be $k$. Also, let $f_1^A$, $f_2^A$,…, $f_k^A$ denote network flows through different edges of cut-set $\boldsymbol{K}$ for a valid graph-based flow solution $A$, and $f_1^B$, $f_2^B$,…, $f_k^B$ denote the network flows through the same edges for a valid graph-based flow solution $B$. Then, by the law of conservation of energy, total power transfer across cut-set $\boldsymbol{K}$ for each of the flow solutions $A$ and $B$ must be equal to $\Delta P_1 = -\Delta P_2$, i.e.,

$$\sum_{l=1}^{k} f_l^A = \sum_{l=1}^{k} f_l^B = \Delta P_1 = -\Delta P_2\,, \quad \forall e_l \in \boldsymbol{K} \tag{3}$$

The validity of (3) is illustrated through three different base-case network flow solutions shown in Figure 2. Cases 1 and 2 denote two valid flow graphs obtained using graph theory while Case 3 depicts a flow graph obtained using a DC power flow solution. Table I shows that for the cut-set $\boldsymbol{K}$ ={4-1, 9-2, 9-3}, even though the flows through the individual edges of the cut-set are different, the net power transfer across cut-set $\boldsymbol{K}$ is equal to 380.86 MW for all three cases. Note that the flow limit of each edge in Figure 2 is 300 MW.

**Table I: Power transfer across cut-set $\boldsymbol{K}$ for three different flow solutions of Figure 2**

| Edges in K | Case 1: Flow (MW) | Case 2: Flow (MW) | Case 3: Flow (MW) |
|---|---|---|---|
| 4-1 | 208 | 35.86 | 172.51 |
| 9-2 | 0 | 72.86 | 121.96 |
| 9-3 | 172.86 | 272.14 | 86.39 |
| **Power transfer** | 380.86 | 380.86 | 380.86 |

## IV. Graph Theory Based Network Analysis Tool

As described in Section II, a transmission line or transformer will be considered a special asset if the power flowing through it cannot be rerouted via the set of its indirect paths. For each such special asset, the graph theory-based network analysis tool finds the limiting critical cut-set as described below.

### A. *Graph theory-based feasibility test (FT)*

The graph theory-based feasibility test (FT) described in **Algorithm II** below examines all the transmission assets to quickly identify the set of special assets and the limiting critical cut-set corresponding to each special asset. That is, if $f_l$ units of power flows through edge $e_l$ from $v_l^F$ to $v_l^T$, **Algorithm II** will first identify if $e_l$ is a special asset. If $e_l$ does turn out to be a special asset, then **Algorithm II** will identify the associated limiting critical cut-set, $\boldsymbol{K}_{crit}$, and the power transfer margin associated with $\boldsymbol{K}_{crit}$, denoted by $T_l$.

Although there can be multiple saturated cut-sets associated with a special asset, **Algorithm II** is able to identify the limiting critical cut-set because it is the first one to get saturated in Step (iv). Consider the system shown in Figure 2 once more. When edge 4-1 is examined by the FT, with respect to any of the three flow graphs, following observation is made: edge 4-1 is a special asset as it fails FT, and is associated with a limiting critical cut-set containing edges 4-1 and 6-7 (i.e., $\boldsymbol{K}_{crit}$={4-1,6-7}) and $T_l$= -35.86 MW. The implication of the above statement is explained with the help of Figure 3(a), Figure 3(b), and Figure 3(c) which depict the power transfer across cut-set $\boldsymbol{K}_{crit}$ for the three different flow graphs of Figure 2.

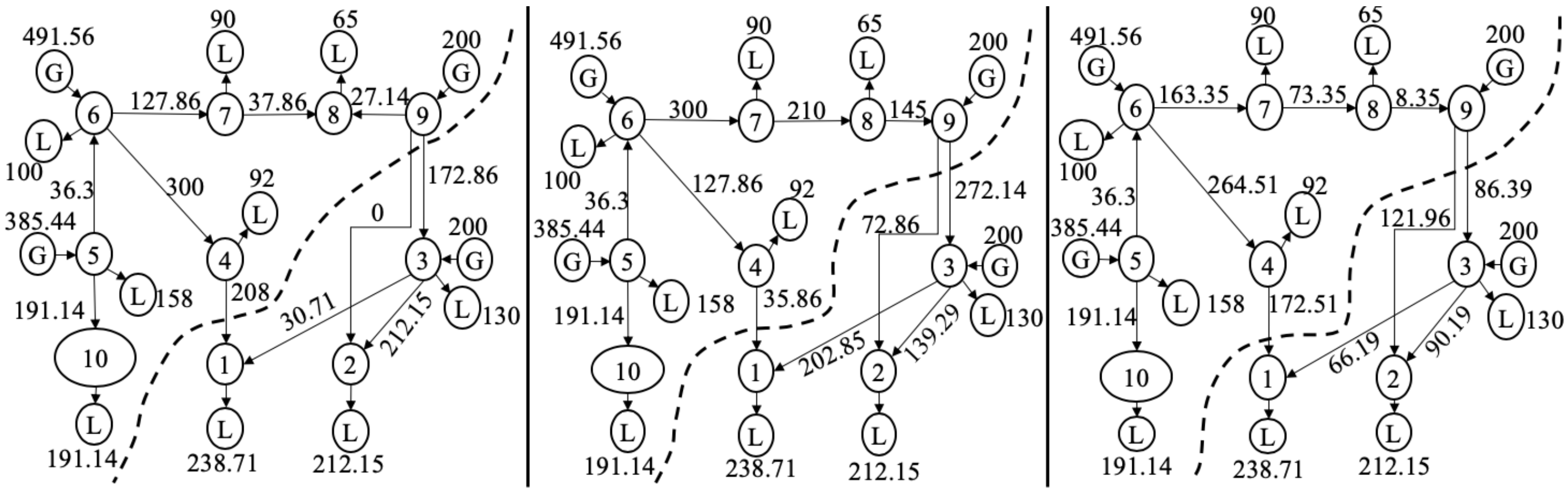

(a) Case 1 (b) Case 2 (c) Case 3

Figure 2: Three valid graph theory-based flow graphs for the same system. “G” and “L” denote generation and load respectively; dotted line denotes a cut-set.

**Algorithm II**: Graph theory-based feasibility test (FT)

i. Define $\mathcal{C}'(\boldsymbol{V},\boldsymbol{E}) = \mathcal{C}(\boldsymbol{V},\boldsymbol{E})$. Remove edge $e_l$ from $\mathcal{C}'$. Initialize a variable $TC_l$ to zero (i.e., $TC_l \coloneqq 0$).
ii. Search $\mathcal{C}'$ to obtain the shortest unsaturated path $\boldsymbol{P}$ from $v_l^F$ to $v_l^T$ using breadth first search (BFS) [37]; path $\boldsymbol{P}$ is considered unsaturated if it has capacity to reroute additional flow.
iii. Find the maximum extra flow, $C_P$, that can be rerouted through path $\boldsymbol{P}$ from $v_l^F$ to $v_l^T$ .
iv. Update $TC_l$ as $TC_l \coloneqq TC_l + C_P$, and the weights of $\mathcal{C}'$ as per (1); note that this step saturates path $\boldsymbol{P}$ in $\mathcal{C}'$.
v. Repeat Steps (ii) through (iv) until there exists no unsaturated path in $\mathcal{C}'$ from $v_l^F$ to $v_l^T$.
vi. Due to outage of $e_l$, compute the transfer margin, $T_l$, as: $T_l = TC_l - f_l$. If $T_l$ for $e_l$ is negative, $e_l$ is a special asset.
vii. To identify $\boldsymbol{K}_{crit}$, traverse the saturated graph $\mathcal{C}'$ from $v_l^F$ to $v_l^T$. All the vertices that can be reached from $v_l^F$ without traversing a saturated edge are grouped into cluster $\boldsymbol{C}_1$. Similarly, the vertices that cannot be reached from $v_l^F$ without traversing a saturated edge are grouped into cluster $\boldsymbol{C}_2$. Cut-set $\boldsymbol{K}_{crit}$ contains the edges whose one end is in $\boldsymbol{C}_1$ and the other end is in $\boldsymbol{C}_2$.

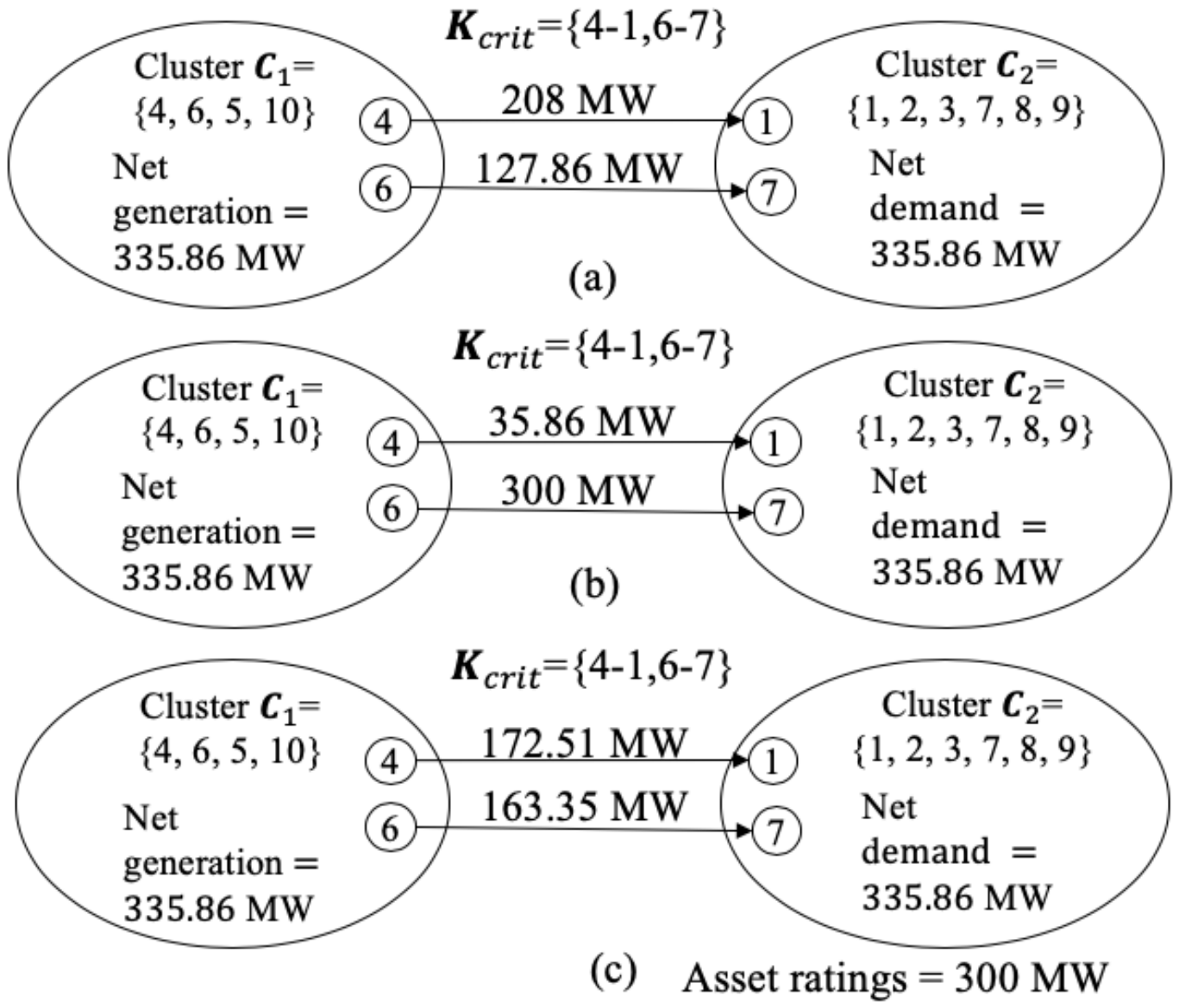

Figure 3: (a) Power transfer across cut-set $\boldsymbol{K}_{crit}$ for flow graph of Figure 2(a), (b) Power transfer across cut-set $\boldsymbol{K}_{crit}$ for flow graph of Figure 2(b), and (c) Power transfer across cut-set $\boldsymbol{K}_{crit}$ for flow graph of Figure 2(c)

From Figure 3 it is clear that although the individual flows on different edges of the cut-set are different, FT finds that, for all three flow graphs, if the edge 4-1 is lost, the cut-set $\boldsymbol{K}_{crit}$ will have a power transfer capability shortage of 35.86 MW from cluster $\boldsymbol{C}_1$ to cluster $\boldsymbol{C}_2$. For example, in Figure 3(a), when edge 4-1 is lost, the flow in edge 6-7 becomes (208+127.86) MW = 335.86 MW, which exceeds its rating (of 300 MW) by 35.86 MW. In summary, the FT: (a) detects special assets, (b) identifies the limiting critical cut-set associated with each special asset, and (c) computes the power transfer margin across the identified limiting critical cut-set.

### *B. Graph theory-based network flow update scheme (UPS)*

During major power system disturbances, multiple outages can occur in rapid succession. Therefore, the FT results would also change following the outage of an edge. To identify the set of special assets following an outage, it is important to first update the graph theory-based network flows to account for the outage of any edge. The advantage of graph theory-based flows is that rerouting of the flow upon the loss of an edge can be achieved extremely fast. The technique of updating the flow graph $\mathcal{F}(\boldsymbol{V},\boldsymbol{E})$ and latent capacity graph $\mathcal{C}(\boldsymbol{V},\boldsymbol{E})$ when edge $e_l$ suffers an outage is done in accordance with **Algorithm III**, which describes the graph theory-based update scheme (UPS). The UPS for the outage of an edge is explained with the help of Figure 4. A flow graph obtained from graph theory-based network flow algorithm is depicted in Figure 4(a). The update of the network flows when the edge 5-6 goes out is shown in Figure 4(b). The UPS simply reroutes 25 MW of flow through path 5-4-1-6 to create an updated network flow solution.

**Algorithm III**: Graph theory-based update scheme (UPS)

i. Let, the flow to be rerouted be given by $F = f_l$, where $f_l$ refers to the flow through edge $e_l$ from vertex $v_l^F$ to $v_l^T$.
ii. Remove edge $e_l$ from $\mathcal{F}(\boldsymbol{V},\boldsymbol{E})$ and $\mathcal{C}(\boldsymbol{V},\boldsymbol{E})$.
iii. Search $\mathcal{C}$ to obtain the shortest unsaturated path $\boldsymbol{P}$ from $v_l^F$ to $v_l^T$ using breadth first search (BFS) [37].
iv. Find the maximum extra power, $C_P$, that can be rerouted through path $\boldsymbol{P}$.
v. If $F > C_P$, inject $C_P$ units of flow through path $\boldsymbol{P}$ and update $F$ as $F \coloneqq F - C_P$. If $F \le C_P$, inject $F$ units of flow through path $\boldsymbol{P}$ and set $F \coloneqq 0$. Update the weights of $\mathcal{F}$ and $\mathcal{C}$ accordingly.
vi. Repeat Steps (ii) through (v) until $F = 0$.

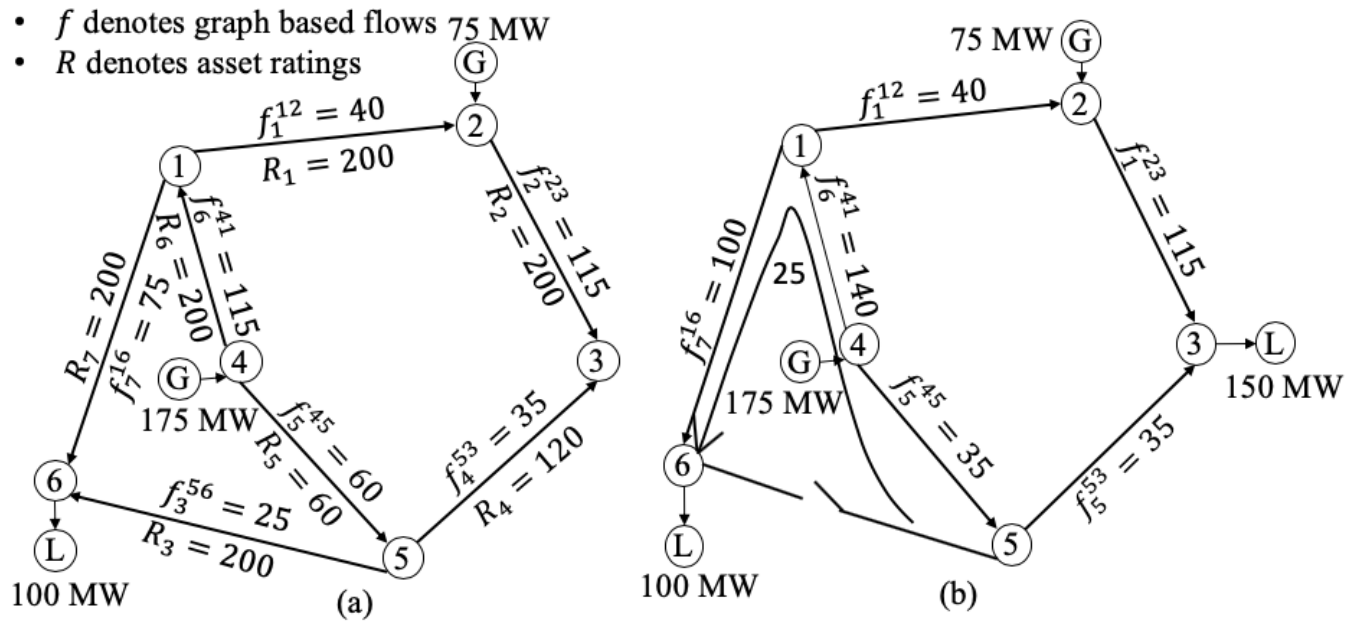

Figure 4: (a) A flow graph obtained from graph-based network flow algorithm, (b) Update scheme (UPS) of network flow solution for the outage of edge 5-6

### *C. Shortlisting assets (SA) scheme for feasibility test (FT)*

In the base-case scenario when the flow graph is built for the first time all transmission assets would be investigated by the FT. However, in the event of the outage of an edge, when UPS gives an updated flow graph, it is not necessary to test all the assets by the FT once again to identify the special assets. By intelligently exploiting the information provided by FT in the base-case scenario and using the UPS to reroute the flow for the edge that is out, the FT can be performed on only a subset of the assets to evaluate the impact of a second contingency. This is explained through Figure 5.

Let it be known from the base-case FT that flow through edge $e_m$ can be rerouted through path $\boldsymbol{P}_1$, while the loss of edge $e_l$ alters flow through path $\boldsymbol{P}_2$. Then, in Figure 5(a), when $e_l$ goes out, the flow through $e_l$ is rerouted through $\boldsymbol{P}_2$ by UPS. Now, as $\boldsymbol{P}_1$ and $\boldsymbol{P}_2$ do not involve common edges, the rerouting of power through $\boldsymbol{P}_2$ by UPS does not modify the flows through $\boldsymbol{P}_1$; therefore, FT need not be repeated for $e_m$. However, if $\boldsymbol{P}_1$ and $\boldsymbol{P}_2$ have common edges, as seen in Figure 5(b); i.e., rerouting of the flow of $e_l$ affects the flow through $\boldsymbol{P}_1$, then $e_m$ must be examined by FT once again after the outage of $e_l$. This rationale of screening the assets to be examined by FT in the event of an outage is called the shortlisting asset (SA) scheme. It will be shown in Section V that the usage of the SA scheme significantly reduces the computation time.

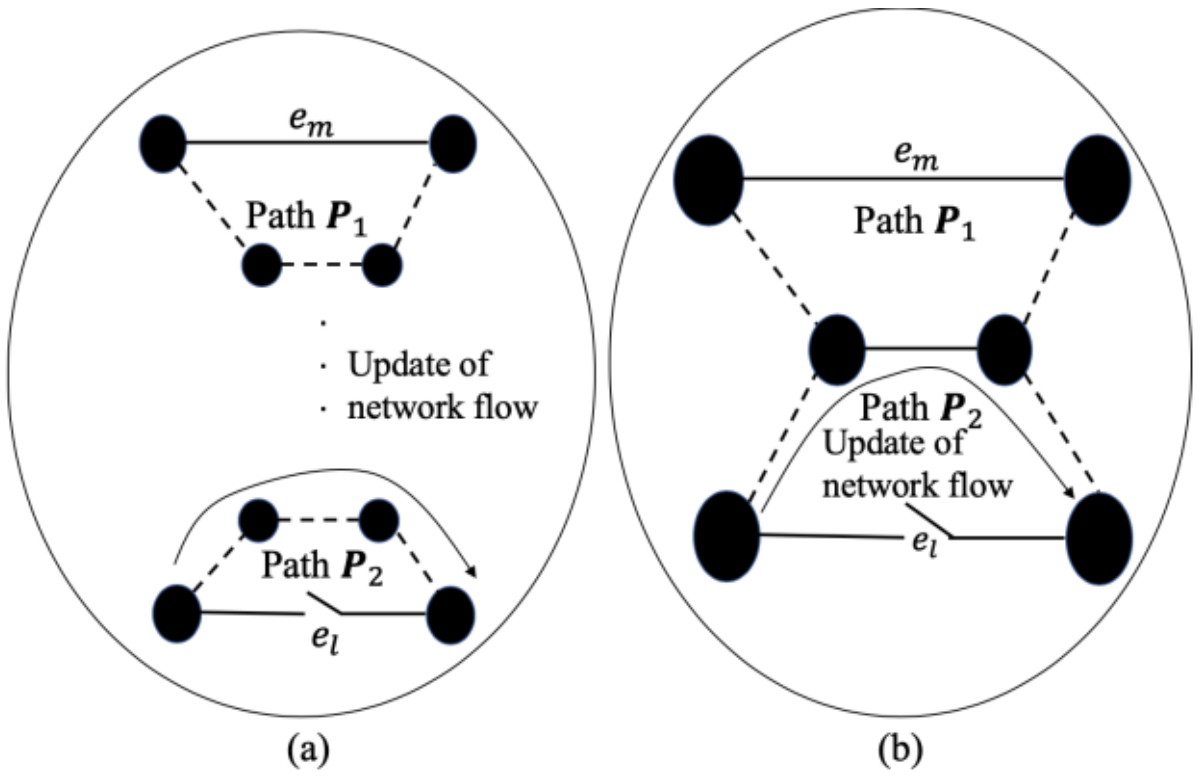

Figure 5: (a) Rerouting the flow on edge $e_l$ does not involve any edge of the indirect paths of edge $e_m$, and (b) Rerouting the flow on edge $e_l$ involves some edges of the indirect paths of edge $e_m$

### D. Graph-traversal scheme

The proposed FT and UPS algorithms use BFS [37] scheme to traverse the graph. BFS, in comparison to depth first search (DFS), has the advantage that it starts at a source vertex and explores all the neighboring vertices at present depth before moving on to the vertices at the next depth. Once the sink vertex is reached the algorithm stops. When BFS is used to traverse the graph to reach a sink from a given source, the path traced by BFS is already the shortest path (if there was a shorter path, BFS would have found it earlier). Both the graph theory-based FT and UPS scan through the set of indirect paths associated with any edge. However, there could be many indirect paths associated with an edge. The unique search properties of FT and UPS that facilitates real-time identification of limiting critical cut-sets and rerouting of the network flow, respectively, are discussed below:

*1) Graph traversal during feasibility test (FT)*:

In each iteration of the FT, saturation of an indirect path occurs as described in Step (iv) of **Algorithm II**. This will occur within a small number of iterations for a power network because many of its edges are common to multiple indirect paths. Therefore, the limiting critical cut-set, $\boldsymbol{K}_{crit}$, can be identified without requiring the BFS scheme to scan through the set of all indirect paths.

*2) Graph theory-based update scheme (UPS)*:

Based on the same rationale explained above, for rerouting the flow through any edge of the power network, **Algorithm III** will only utilize a small subset of indirect paths.

Apart from BFS, other commonly used graph search methods for finding the shortest path between a source-sink pair are Bellman-Ford algorithm [38], and Dijkstra algorithm [39]. If |**E**| denotes the total number of edges, and |**V**| denotes the total number vertices, the time-complexity of the Bellman-Ford algorithm is O(|**E**||**V**|) [40]. The time-complexity of Dijkstra algorithm implemented using binary heap is $O(|\boldsymbol{E}| + |\boldsymbol{V}|log|\boldsymbol{V}|)$ [41]. Lastly, the time-complexity of the BFS algorithm is $O(|\boldsymbol{E}| + |\boldsymbol{V}|$ [42], which is the best among the three shortest-path graph traversal techniques. Therefore, we have used the BFS graph traversal scheme to design the proposed algorithms (FT and UPS) to determine if contingencies create saturated cut-sets.

## V. Results

### A. IEEE 118-bus system

The utility of the proposed algorithm for enhanced situational awareness is explained with a case-study on IEEE-118 bus system. Due to a hurricane, let the following transmission asset outages occur one after another: 15-33, 19-34, 37-38, 49-66, and 47-69 (marked $O_1$ through $O_5$ in Figure 6). From Figure 6 and Table II, following information is obtained when the algorithm is applied as outages manifest:

1) *Base-case:* In the base-case scenario, the asset 26-30 fails the graph theory-based FT and is classified as a special asset. The loss of 26-30 would saturate the limiting critical cut-set $\boldsymbol{K}_{crit}^{0}$ by a margin of -77 MW, i.e., $T_l^0$= -77 MW.
2) *1st Outage*: When 15-33 is lost, no additional special assets are identified.
3) *2nd Outage*: When 19-34 is lost, no additional special assets are identified.
4) *3rd Outage*: When 37-38 is lost, the asset 42-49 fails the FT and is classified as a special asset. The loss of 42-49 would saturate the limiting critical cut-set $\boldsymbol{K}_{crit}^{3}$ by a margin of -186 MW, i.e., $T_l^3$= -186 MW.
5) *4th Outage*: When 49-66 is lost, no additional special assets are identified.
6) *5th Outage*: When 47-69 is lost, the assets 59-56, 63-59, 63-64, and 64-65 are classified as special assets. The loss of these four assets would saturate the limiting critical cut-sets, $\boldsymbol{K}_{crit}^{5a}$, $\boldsymbol{K}_{crit}^{5b}$, $\boldsymbol{K}_{crit}^{5c}$, and $\boldsymbol{K}_{crit}^{5d}$, by margins of -64, -191, -191, and -219 MW, respectively (i.e., $T_l^{5a}$= -64 MW, $T_l^{5b}$= -191 MW, $T_l^{5c}$= -191 MW, $T_l^{5d}$= -219 MW).

The value of the information obtained above can be realized by considering the following scenario: after the occurrence of the fifth outage, the proposed algorithm would inform the power system operator that if any of the four assets identified in the last row, second column of Table II is lost next (as the 6th outage), the corresponding cut-set identified in the third column would be saturated by the margin mentioned in the fourth column. If this anticipated overload is to be avoided, the operator must preemptively reduce the power flowing through the identified cut-set by *at least* the amount mentioned in the last column. Thus, the proposed network analysis tool is an enhanced power system connectivity monitoring scheme that improves the power system operators' situational awareness by augmenting their visualization in real-time.

**Table II: Identification of limiting critical cut-sets**

| Event | New special asset | Limiting critical cut-set | Transfer margin (MW) |
|---|---|---|---|
| Base-case | 26-30 | $\boldsymbol{K}_{crit}^{0}$ = {26-30,25-27,25-23} | $\boldsymbol{T}_l^0$=-77 |
| Outage 1 (15-33) | - | - | - |
| Outage 2 (19-34) | - | - | - |
| Outage 3 (37-38) | 42-49 | $\boldsymbol{K}_{crit}^{3}$= {42-49,44-45} | $\boldsymbol{T}_l^3$= -186 |
| Outage 4 (49-66) | - | - | - |
| Outage 5 (47-69) | 59-56 | $\boldsymbol{K}_{crit}^{5a}$= {59-56,59-54,59-55,69-49} | $\boldsymbol{T}_l^{5a}$= -64 |
| | 63-59 | $\boldsymbol{K}_{crit}^{5b}$= {63-59,61-59,60-59,69-49} | $\boldsymbol{T}_l^{5b}$= -191 |
| | 63-64 | $\boldsymbol{K}_{crit}^{5c}$= {63-64,61-59,60-59,69-49} | $\boldsymbol{T}_l^{5c}$= -191 |
| | 64-65 | $\boldsymbol{K}_{crit}^{5d}$= {64-65,66-62,66-67,69-49} | $\boldsymbol{T}_l^{5d}$= -219 |

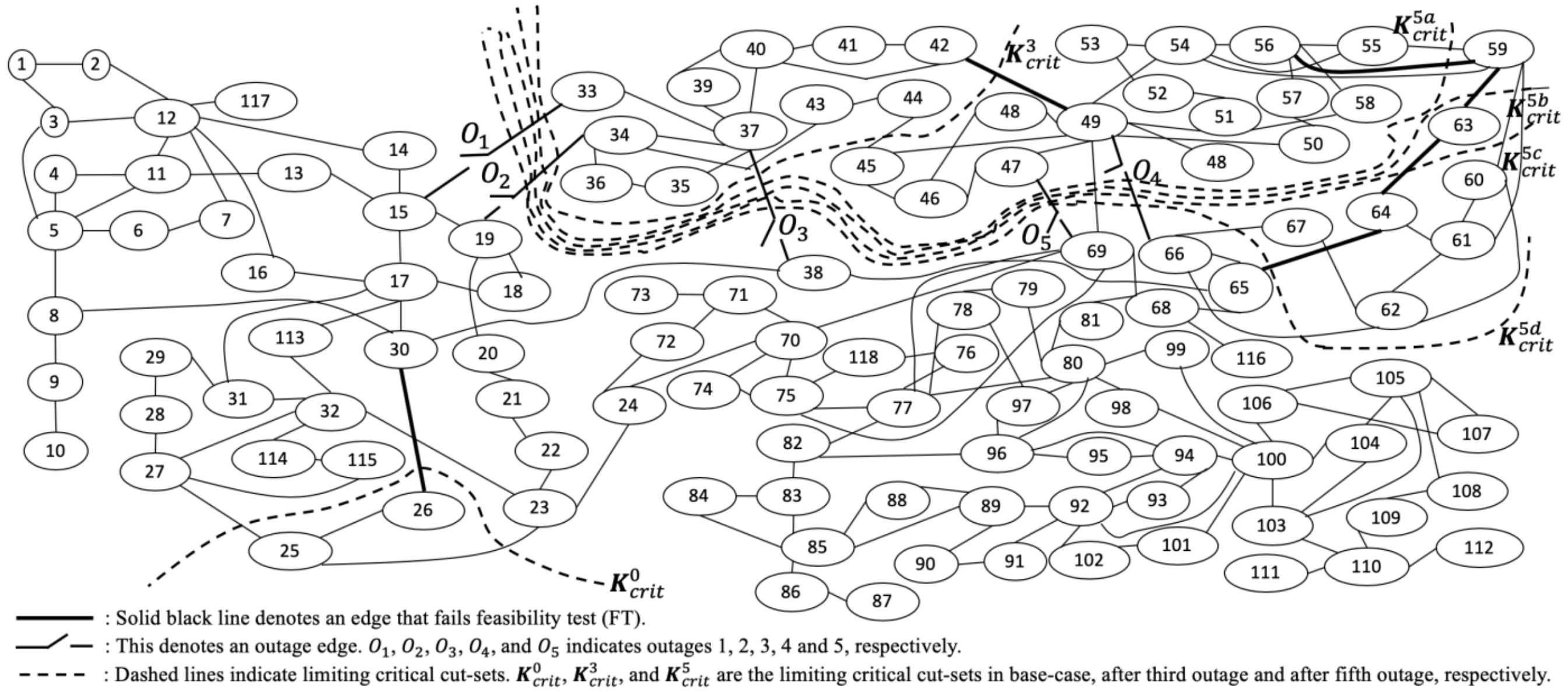

Figure 6: Real time identification of limiting critical cut-sets on the IEEE 118-bus test system

## *B. 17,941 bus model of Western Interconnection (WI)*

This section presents the results obtained using a 17,941-bus model of WI. Sub-section 1) presents some statistics of graph theory-based FT and UPS which highlight the computational advantage of the proposed technique. Sub-section 2) describes how the proposed network analysis scheme provides useful information when a sequence of outages occurs in this system.

### *1) Computational efficiency of the graph theory-based FT:*

It takes 6 min to run *N*-1 FT for this system. Conversely, the time required to run a single DC power flow is 0.12 s. As there are 22,091 transmission assets in the system, time required to run a DC power flow for the outage of every asset would require solving 22,091 DC power flows, which would approximately require 0.12×22,091 s ≅ 44 min. Therefore, for this system, performing *N*-1 FT is approximately 7 times faster compared to *N*-1 DC CA. For a valid comparison, the simulations were done on the same computer (Core i7, 3.60 GHz CPU processor with 16 GB RAM).

When any edge $e_l$ is examined by the graph theory-based FT, the indirect paths of $e_l$ are traversed by BFS. However, the saturation of the set of indirect paths may occur after a small number of indirect paths are traversed by the graph theory-based FT. Moreover, since BFS always identifies the shortest path from the source to the sink, the number of edges contained in an indirect path would be relatively small. For every non-radial edge of this system, the number of indirect paths required to saturate the graph and the maximum number of edges contained in an indirect path is computed. The statistics of FT is summarized in Figure 7(a) and Figure 7(b).

Figure 7(a) plots the histogram for the number of indirect paths used by BFS to saturate the latent capacity graph. The largest number of indirect paths required was 58. Figure 7(b) plots the histogram of maximum number of edges contained in an indirect path traced by BFS; the maximum was 111. Thus, the histogram plots demonstrate that the graph theory-based FT essentially uses a small subgraph to detect the saturation of a cut-set; this is the fundamental reason why the graph theory-based FT is computationally so efficient.

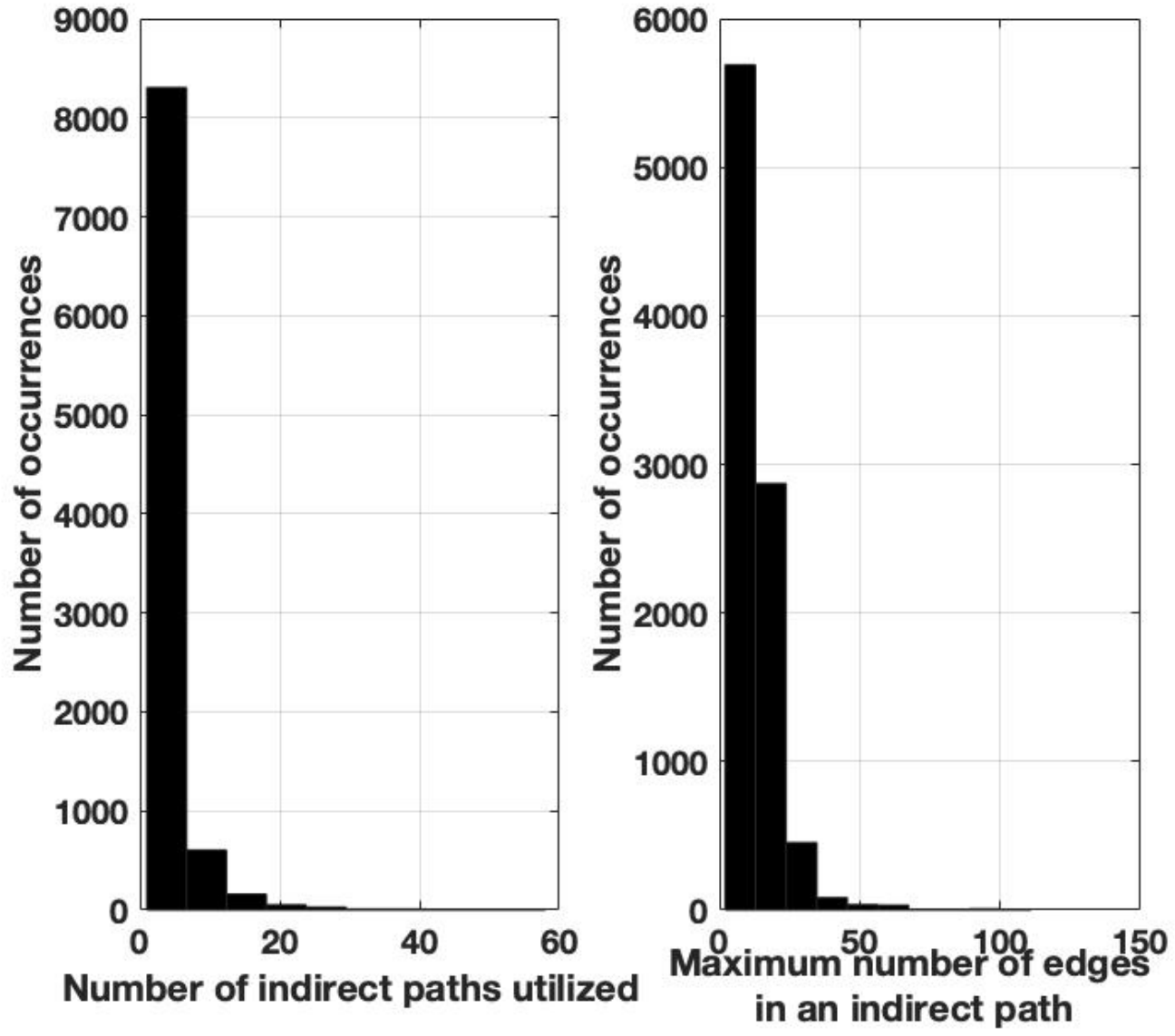

Figure 7: (a) Histogram of number of indirect paths traversed by the graph theory-based FT, and (b) Histogram of maximum number of edges contained in an indirect path

### *2) A case study on Western Interconnection (WI):*

This sub-section demonstrates the usefulness and scalability of the proposed approach through a *N*-1-1 event analysis of this system. The loss of 500 kV Hassayampa-North Gila (H-NG) transmission line was the first event, while the second event was the loss of 230/92 kV Coachella Valley transformers. Before the analysis was done for the outage of the events, it took approximately 0.5 s to build the flow graph and the latent capacity graph for the base-case. As mentioned earlier, it takes approximately 6 min to run FT on all transmission assets in the base-case. Whether events 1 and 2 resulted in any additional special asset was investigated as follows:

***Event 1***: Once the 500 kV H-NG transmission line was lost, graph theory-based UPS took only 0.20 s to reroute the flow to obtain a new flow graph. The SA scheme took 0.06 s to identify 271 edges that were to be examined by FT for this new graph. Time required by FT to examine all the 271 edges for an outage was 32 s. Among the 271 edges, 4 edges failed FT and were classified as special assets as shown in Table III. For the four special assets, the FT found the corresponding limiting critical cut-set, $\boldsymbol{K_{crit}}$; $\lvert \boldsymbol{K_{crit}} \rvert$ in Table III denotes the number of edges contained in $\boldsymbol{K_{crit}}$. Moreover, FT provided information regarding the impact of the loss of a special asset on the associated limiting critical cut-set. For example, if the transmission corridor 936-1192 is lost next, the limiting critical cut-set would be saturated by a margin of 441 MW. The total time required to perform this network analysis and identify all the limiting critical cut-sets after the outage of H-NG was 32.26 s (i.e., total time taken by UPS, SA, and FT). On the other hand, if FT were to be run on all transmission assets (as was done in the base-case), the time required would be 6 min. Therefore, intelligently performing FT on a shortlisted set of edges reduced the computation time from 6 min to 32.26 s.

**Table III: Application of the graph theory-based network analysis in Western Interconnection (WI)**

| Events | Time required by UPS | SA for FT | | FT on shortlisted assets | | | | Total time |
|---|---|---|---|---|---|---|---|---|
| | | #Edges | Time | New special assets | $\lvert K_{crit} \rvert$ | $T_m$ | Time | |
| Line outage: Hassayampa-North Gila | 0.20 s | 271 | 0.06 s | 936-1192 | 57 | −441 MW | 32 s | (0.20+0.06+32) =32.26 s |
| | | | | 1192-1217 | 49 | −1258 MW | | |
| | | | | 2873-2902 | 18 | −419 MW | | |
| | | | | 2902-2903 | 21 | −309 MW | | |
| Transformer outage: Coachella Valley | 0.06 s | 82 | 0.07 s | 2416-2488 | 8 | −35.35 MW | 10 s | (0.06+0.07+10) = 10.13 s |
| | | | | 2421-2487 | 2 | −2 MW | | |
| | | | | 2421-3293 | 2 | −2 MW | | |
| | | | | 2438-2606 | 5 | −55 MW | | |
| | | | | 2487-2488 | 8 | −35 MW | | |
| | | | | 2712-2878 | 9 | −35 MW | | |

***Event 2***: When 230/92 kV Coachella Valley transformers are tripped, the UPS took only 0.06 s to obtain the updated network flow solution. Time required by the SA scheme to shortlist the edges to be examined by FT was 0.07 s; 82 new edges were shortlisted. Time required by FT to examine all the 82 shortlisted edges was 10 s. Among the 82 edges examined, 10 edges failed FT and were classified as special assets (see Table III). Total time required to identify the set of special assets after the outage of Coachella Valley transformers was 10.13 s. Conversely, performing a DC CA for all transmission assets would have required 44 min, or about 260 times longer than the proposed graph theory-based approach.

## VI. Discussions

### A. Practical utility of the proposed algorithm

After the 2011 U.S. Southwest blackout, the FERC [3] reported the following finding: "Affected TOPs (transmission operators) have limited visibility outside their systems, typically monitoring only one external bus. As a result, they lack adequate situational awareness of external contingencies that could impact their systems. They also may not fully understand how internal contingencies could affect SOLs (system operating limits) in their neighbors' systems." The recommendation of FERC to TOPs was to "review their real-time monitoring tools, such as state estimator and RTCA, to ensure that such tools represent critical facilities needed for the reliable operation of the BPS (bulk power system)".

Now, modeling all "critical facilities" over a large area (across different utilities) could significantly increase the number of contingencies to be evaluated by RTCA, which would then increase the solution time considerably [4], [11]. In this regard, the *ability of the proposed algorithm to analyze the effects of any outage on very large systems and provide meaningful quantifiable information in a matter of seconds gives it a distinct advantage*. Moreover, the special assets detected by the FT can be suitable candidates for detailed analysis by a more precise CA tool. Thus, the proposed research can complement real-time operations by extending an operator's visibility to external contingencies, while alleviating the associated computational burdens.

### B. Proposed method is not guaranteed to detect all contingencies that result in post-contingency branch overloads

As per the FT when all the indirect paths do not have sufficient capacity to reroute the power flowing through an edge, it implies that it would inevitably result in post-contingency branch overloads. However, the converse is not true. This is illustrated using the test system shown in Figure 8, and the corresponding flows shown in Figure 9 and Figure 10.

Figure 9(a) presents a DC power flow solution, when 100 MW of power is injected at bus 1, and 100 MW is withdrawn at bus 2 (Scenario 1). The numbers in non-bold fonts indicate flows, while the numbers in bold font denote line ratings. The proposed FT algorithm identifies edge 1-2 as a special asset because the indirect paths of edge 1-2 do not have sufficient capacity to reroute the flow through the direct path, namely, edge 1-2. A post-contingency DC power flow shown in Figure 9(b) validates that such an outage results in overloads along Indirect path 1. Figure 10(a) presents a DC power flow solution, when 85 MW of power is injected at bus 1, and 85 MW is withdrawn at bus 2 (Scenario 2). In this scenario, the proposed FT algorithm does not identify edge 1-2 as a special asset because the set of indirect paths have sufficient capacity to reroute the flow of the direct path. However, a post-contingency DC power flow solution shown in Figure 10(b) indicates that the Indirect path 1 is still overloaded, due to lower impedance of Indirect path 1 compared to Indirect path 2.

From this illustration, the following conclusions can be drawn: when the set of indirect paths do not have the capacity to reroute the power flowing through the direct path (see Figure 9), no additional information is required to conclude that there would be a post-contingency overload. The proposed graph theoretic power flow model takes advantage of this observation to identify violations quickly. At the same time, the proposed approach is not able to capture the overload occurring in Figure 10. This is because the graph theory-based network flow algorithm ignores the effects of impedances when creating the flows. Thus, the proposed approach may not detect all possible post-contingency branch overloads.

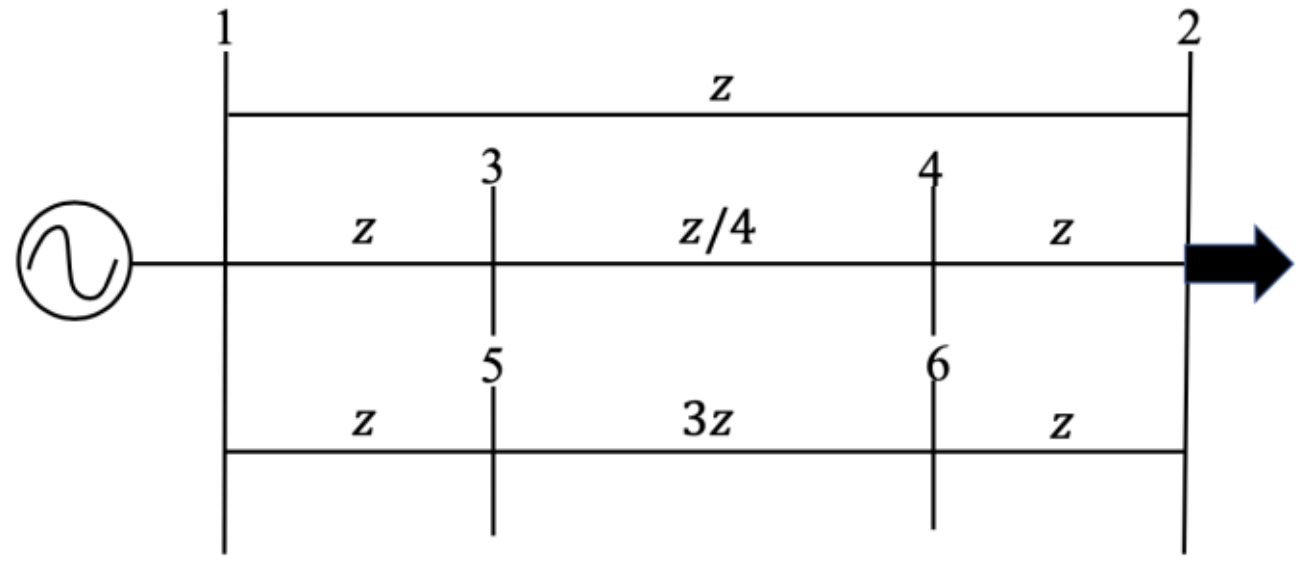

Figure 8: Topology of a sample six-bus power system (branch impedances are represented in terms of a variable $z$)

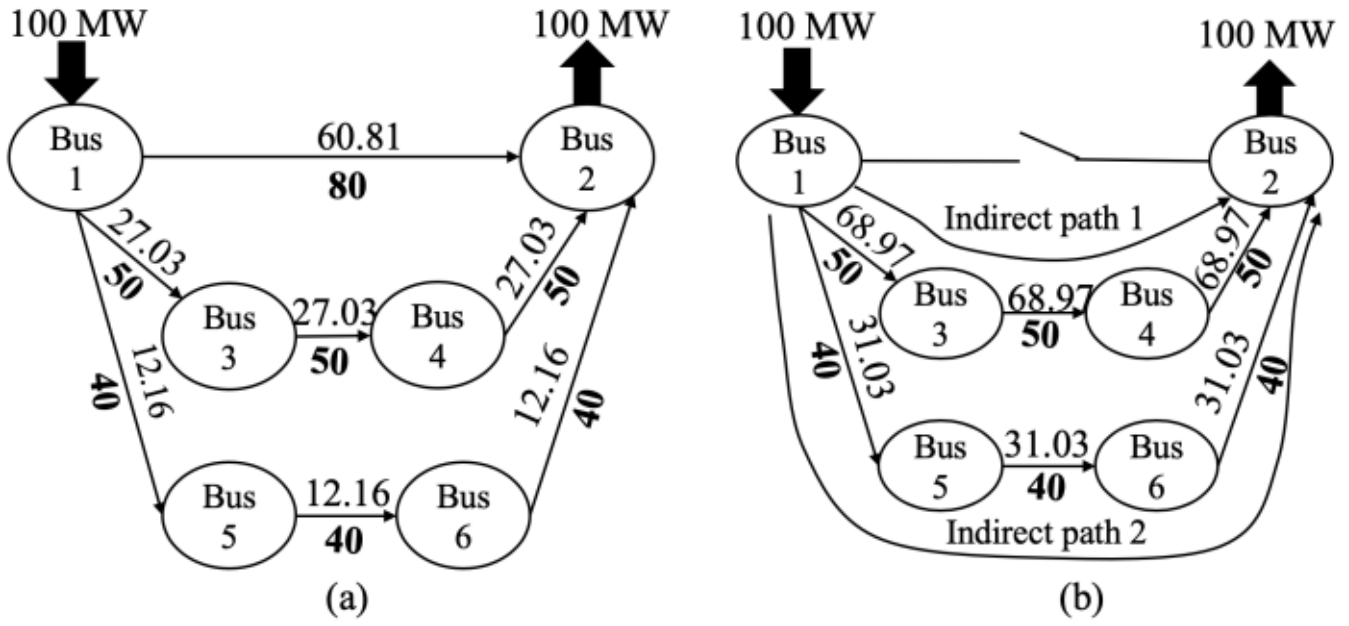

Figure 9: Scenario 1. (a) A DC power flow solution in base-case, and (b) A DC power flow solution for the outage of edge 1-2

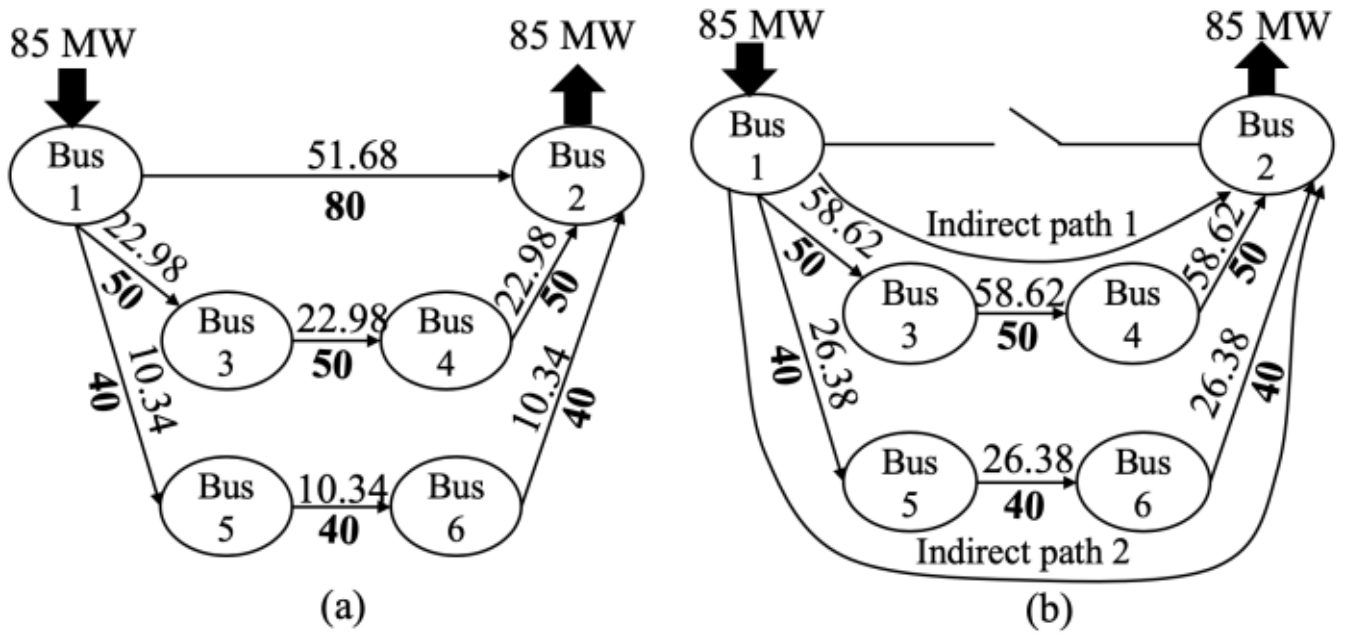

Figure 10: Scenario 2. (a) A DC power flow solution in base-case, and (b) A DC power flow solution for the outage of edge 1-2

### *C. Proposed method is guaranteed to detect all contingencies that saturate a cut-set*

The discussion presented in Section VI.B reveals that the graph theory-based FT is not guaranteed to identify all contingencies that create post-contingency branch overloads. However, *the proposed algorithm does guarantee detection of all contingencies that create a saturated cut-set in the network.* This is explained as follows. Let us examine if the outage of edge $e_l$ of Figure 11 would create a saturated cut-set in the system using the proposed FT. Edge $e_l$ could be associated with multiple cut-sets in the system. With reference to Figure 11 the $i^{th}$ cut-set associated with edge $e_l$ is denoted by

$$\boldsymbol{K}_i = \left\{e_l, e_{l_1}, e_{l_2}, \dots, e_{l_{(k-1)}}\right\} \quad \text{for } 1 \le i \le x \quad (4)$$

where, $k$ is the total number of edges in cut-set $\boldsymbol{K}_i$, and $x$ is the total number of cut-sets associated with edge $e_l$. When the transfer margin, $T_l$, computed by the graph theory-based FT (proposed in Section IV.A) is negative it implies that the outage of edge $e_l$ saturates at least one cut-set, among the $x$ cut-sets that edge $e_l$ is associated with. On the other hand, if the transfer margin, $T_l$, computed by the FT is positive, it implies that the outage of edge $e_l$ does not saturate any of the $x$ cut-sets that it is associated with. Therefore, the graph theory-based FT will not miss a single contingency that would create a saturated cut-set. This is illustrated using the test system shown in Figure 12, and the corresponding flows shown in Figure 13 and Figure 14.

Figure 13 presents a DC power flow solution when the total load and generation in the system is 594 MW (Case 1). The FT algorithm finds that the outage of 3-4 saturates cut-set $\boldsymbol{K}_2$ ={3-4,3-5,1-5} by 31 MW. To validate this inference, the power transfer capability across each of the cut-sets associated with edge 3-4 is enumerated from the DC power flow solution. As shown in Figure 13, edge 3-4 is associated with four cut-sets: $\boldsymbol{K}_1$,$\boldsymbol{K}_2$, $\boldsymbol{K}_3$,$\boldsymbol{K}_4$. The power transfer capabilities across the four cut-sets of the test system when edge 3-4 is lost are summarized in Table IV, where $P_K$, denotes the total flow that is to be transferred across the cut-set, and $R_K$ denotes the total capacity of all the edges belonging to the cut-set (excluding edge 3-4 itself). It is observed that $P_K$ is greater than $R_K$ only for cut-set $\boldsymbol{K}_2$ by 31 MW. This verifies that for Case 1, the outage of edge 3-4 would saturate cut-set $\boldsymbol{K}_2$ by 31 MW.

Figure 14 presents a DC power flow solution when the total load and total generation of the system is 486 MW (Case 2). In this case, the FT algorithm detects that the indirect paths of edge 3-4 have positive transfer margins indicating that they have the capacity to carry additional power, if need be. To validate this observation, the power transfer capability across each cut-set associated with edge 3-4 is enumerated from the DC power flow solution (see Table IV). It is observed that $P_K$ is less than $R_K$ for $\boldsymbol{K}_1$, $\boldsymbol{K}_2$, $\boldsymbol{K}_3$, $\boldsymbol{K}_4$. This proves that for Case 2, outage of edge 3-4 does not saturate any cut-set that is associated with it.

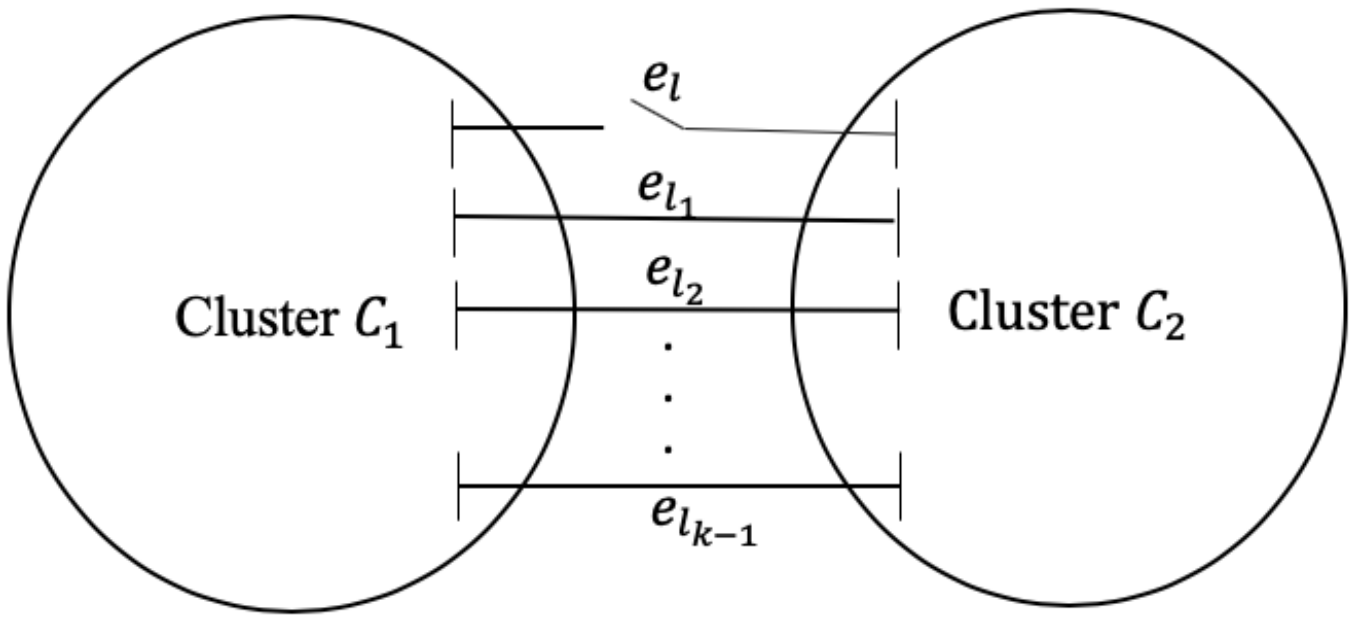

Figure 11: $\boldsymbol{K}_i$ is the $i^{th}$ cut-set (among $x$ cut-sets) associated with edge $e_l$ that separates the network into two disjoint clusters

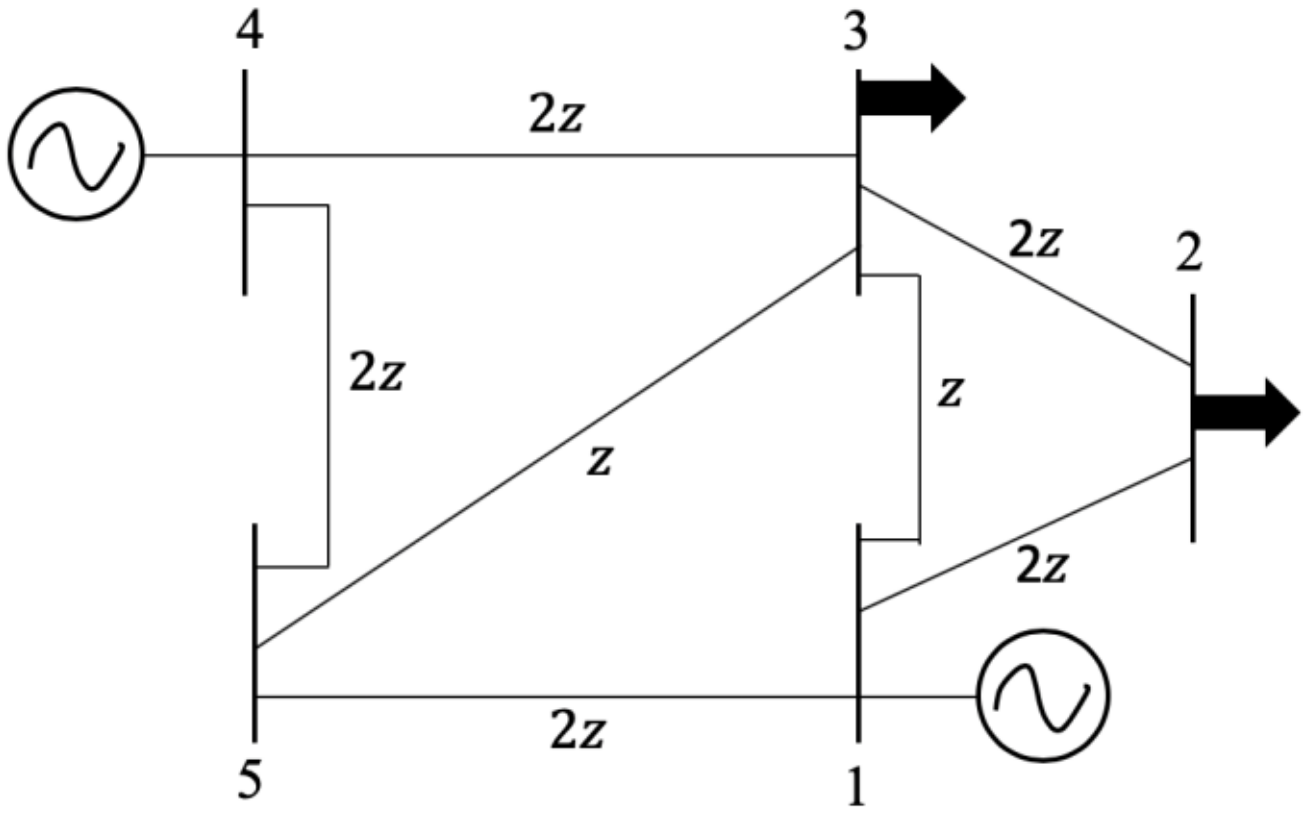

Figure 12: Topology of a sample five-bus power system (branch impedances are represented in terms of a variable $z$)

Furthermore, note that in Figure 13, the power flowing through different edges of the limiting critical cut-set, $\boldsymbol{K}_2$={3-

4,3-5,1-5}, are not in the same direction. This implies that cut-set $\boldsymbol{K}_2$ is not a coherent cut-set (in a coherent cut-set power flows in the same direction in all the edges of the cut-set [27]). Therefore, such types of critical interconnections cannot be detected by the algorithm presented in [27]. It is also important to highlight here that enumerating the power transfer capability across different cut-sets by a DC power flow solution requires previously defining all the cut-sets. On the other hand, the graph theory-based FT can investigate the power transfer capability of different cut-sets without the cut-sets being pre-defined. This is *a unique advantage of the proposed network analysis, because listing all possible cut-sets for a large power network containing thousands of buses especially during extreme event scenarios is not practically feasible.*

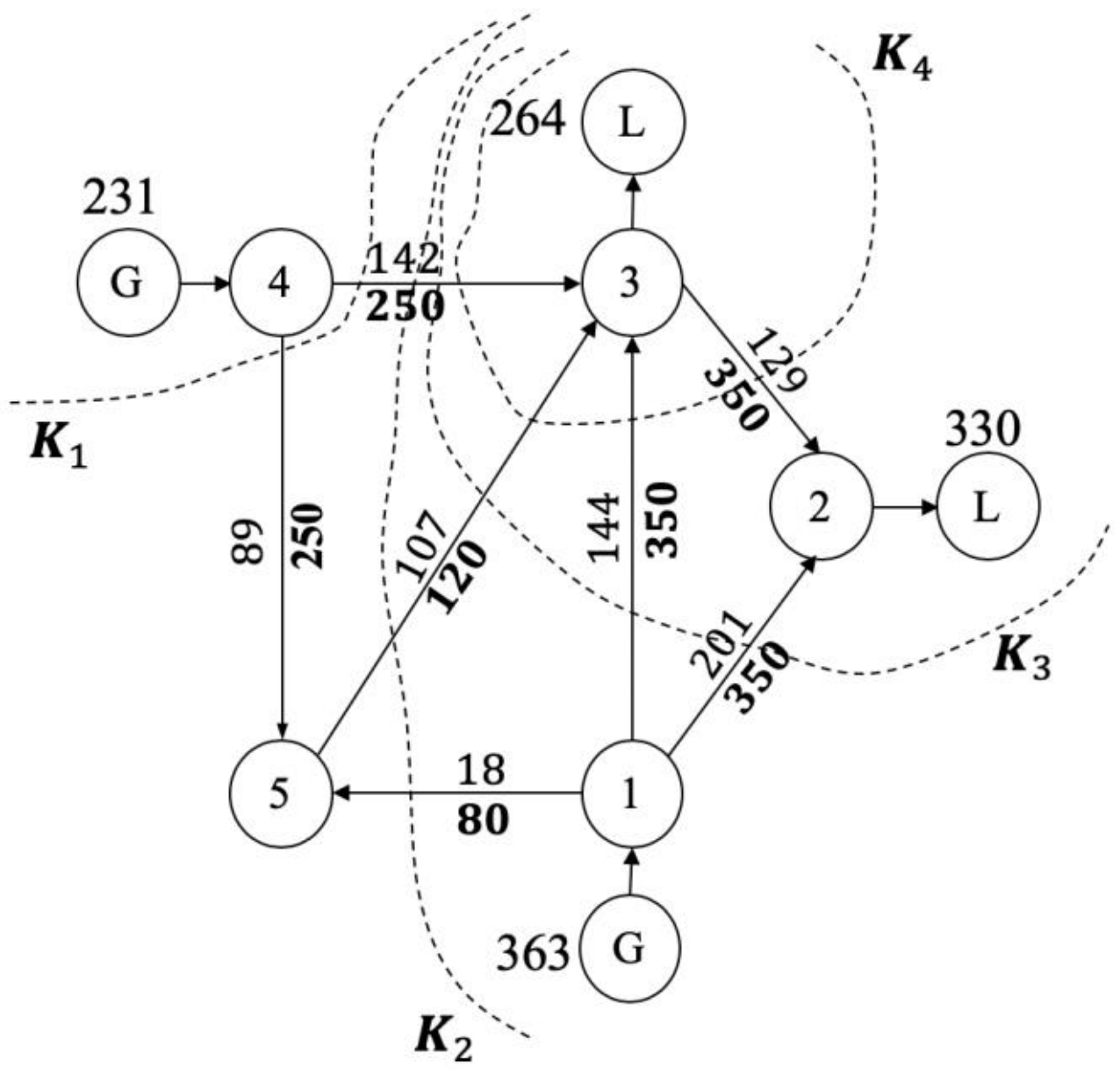

Figure 13: Power transfer across four different cut-sets ($\boldsymbol{K}_1, \boldsymbol{K}_2, \boldsymbol{K}_3, \boldsymbol{K}_4$) associated with edge 3-4 for Case 1

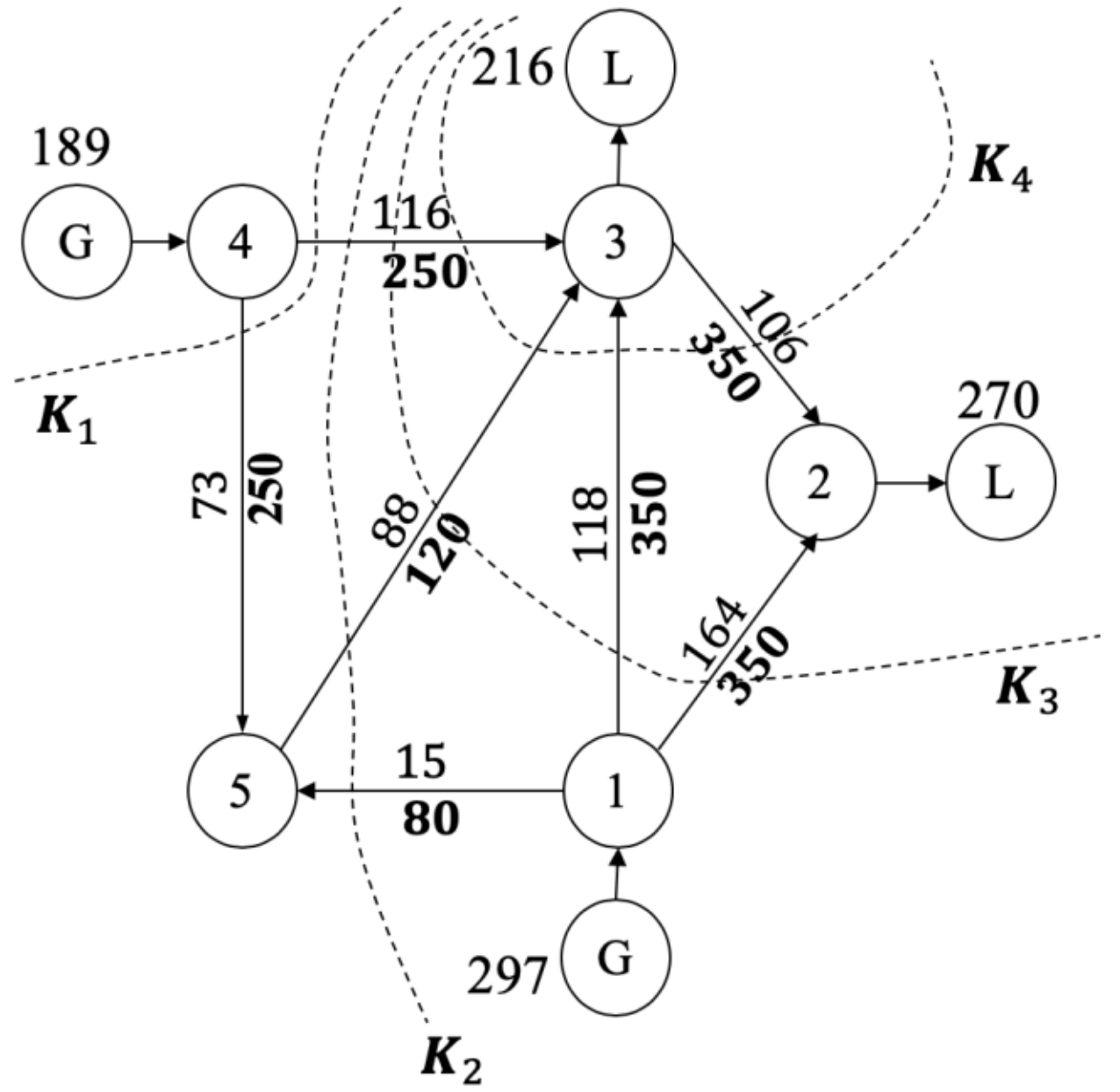

Figure 14: Power transfer across four different cut-sets ($\boldsymbol{K}_1, \boldsymbol{K}_2, \boldsymbol{K}_3, \boldsymbol{K}_4$) associated with edge 3-4 for Case 2

**Table IV: Power transfer capacity across different cut-sets associated with edge 3-4**

| Cut-set | Case 1 | | Case 2 | |
|---|---|---|---|---|
| | Flow ($P_K$) | Capacity ($R_K$) | Flow ($P_K$) | Capacity ($R_K$) |
| $\boldsymbol{K}_1$ | 231 MW | 250 MW | 189 MW | 250 MW |
| $\boldsymbol{K}_2$ | 231 MW | 200 MW | 189 MW | 200 MW |
| $\boldsymbol{K}_3$ | 594 MW | 820 MW | 486 MW | 820 MW |
| $\boldsymbol{K}_4$ | 264 MW | 820 MW | 216 MW | 820 MW |

### *D. Comparison of proposed method with other power system vulnerability assessment techniques*

In this sub-section, the output of the proposed algorithm is compared with those obtained from two power system vulnerability assessment techniques, namely, the metrics developed in [28] and [33], both of which can handle exhaustive *N*-1 evaluation. In [28], an *extended betweenness index* (derived from PTDFs and transmission line limits) was used to rank different contingencies. In [33], a DC power flow based linear sensitivity factor, called *line outage impact factor*, derived from LODFs was used to screen out critical contingencies. The analysis was performed on the IEEE 118-bus system for the same sequence of outages that were described in Table II. In order to validate the severity of the different contingencies, a cascading failure simulation was run in MATCASC [43], a software package linked with MATPOWER that facilitates simulation of cascading failures for any initiating contingency. The amount of load shed at the end of the cascade indicated the severity of the contingency. The results of the comparison are shown in Table V.

**Table V: Ranking of contingencies and cascading analysis on IEEE 118-bus test system**

| Event | Cascading Analysis | | Rank by [28] | Rank by [33] |
|---|---|---|---|---|
| | New contingency | Load shed | | |
| Base-case | 26-30 | 12.20% | 20 | 42 |
| Outage 1 (15-33) | - | - | - | - |
| Outage 2 (19-34) | - | - | - | - |
| Outage 3 (37-38) | 42-49 | 29.87% | 16 | 58 |
| Outage 4 (49-66) | - | - | - | - |
| Outage 5 (47-69) | 64-65 | 28.92% | 6 | 167 |
| | 63-59 | 28.26% | 8 | 70 |
| | 63-64 | 28.26% | 9 | 73 |
| | 56-59 | 25.27% | 15 | 119 |

Column 2 of Table V shows the contingencies identified by MATCASC that result in load shed as the different events manifest in the IEEE 118-bus system. The ranking of these load-shed-causing-contingencies, obtained by the techniques developed in [28] and [33] are provided in Columns 4 and 5, respectively. It can be observed from Table V that the contingencies that actually result in loss of load were not the top ranked contingencies identified by the metrics developed in [28] and [33]. For instance, after the fifth outage, if any of the four new contingencies identified in Column 2 were to occur (as the sixth outage), then it would result in load shedding in excess of 25%. However, none of these four high load-shed-causing-contingencies appeared in the top four ranked contingencies of [28] or [33]. On the other hand, all the load-shed-causing-contingencies were detected as special assets by the proposed algorithm (compare Column 2 of Table V with Column 2 of Table II). This shows the usefulness of the proposed algorithm in detecting critical contingencies.

## VII. Conclusions

This paper presents a new graph theoretic approach for real-time vulnerability assessment resulting in enhanced situational awareness for power system operations. The major contributions of this research are as follows:

1. The most important research finding is that *a relaxed graph theory-based network analysis tool can evaluate if a contingency will create saturated cut-sets in a meshed power system*. The proposed algorithm finds the cut-set which becomes saturated by the largest transfer margin as an impact of the outage. The transfer margin indicates the minimum amount by which the power transfer through the cut-set must be reduced to alleviate the saturation of the cut-set and sustain the impact of the outage.
2. The proposed graph theoretic algorithms (feasibility test, update scheme, and shortlisting asset) reduce the computational time that is required for providing real-time situational awareness. Although the proposed analysis may not detect all types of branch overloads, it is guaranteed to detect all overloaded cut-sets. Fast detection of overloaded cut-sets is important because they represent the "seams or fault-lines across which islanding seems likely" [27].

Using the proposed approach, system operators will also have better preparedness to address the identified violations, long before the set of all possible violations are detected by a more detailed network analysis tool. Hence, the proposed method can be a useful complementary tool to the existing methods of power system vulnerability assessment for real-time operations.

## Acknowledgements

The authors would like to acknowledge Dr. John Undrill from Arizona State University, and Dr. Yingchen Zhang from National Renewable Energy Laboratory for their valuable discussions and help during the course of this research.

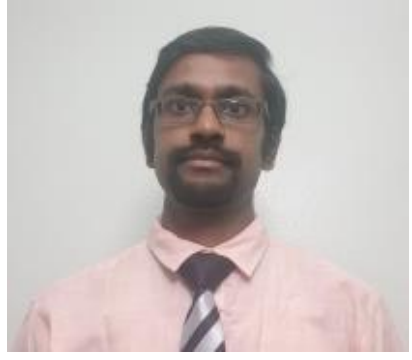

**Reetam Sen Biswas (S’15)** received the B.Tech. degree in electrical engineering from the West Bengal University of Technology, Kolkata, India, in 2016. He received the M.S. degree in electrical engineering from Arizona State University, Tempe, AZ, USA in 2019, where he is currently working towards the Ph.D. degree. His current research interests include power system contingency analysis, cascading failure analysis, state estimation, and power system planning studies with high penetration of renewable energy.

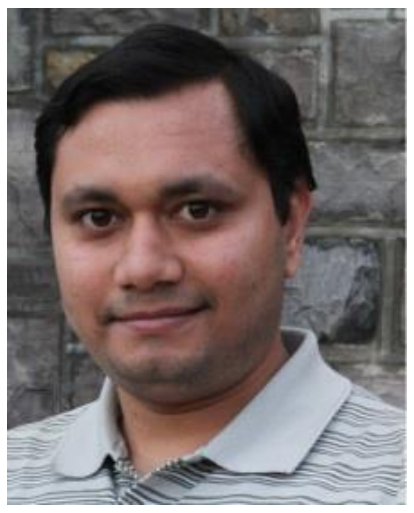

**Anamitra Pal (S’11, M’15, SM’19)** received his Bachelor of Engineering (B.E.) degree (summa) in electrical and electronics engineering from Birla Institute of Technology, Mesra, Ranchi (India) in 2008 and his M.S. and Ph.D. degrees in electrical engineering from Virginia Tech, Blacksburg in 2012 and 2014, respectively. He is now an Assistant Professor in the School of Electrical, Computer, and Energy Engineering at Arizona State University. Previously, from 2014-2016, he was an Applied Electrical and Computer Scientist in the Network Dynamics and Simulation Science Laboratory at the Biocomplexity Institute of Virginia Tech. He is the recipient of the 2018 Young CRITIS Award for his contributions to critical infrastructure resiliency, as well as the 2019 Outstanding IEEE Young Professional Award from the IEEE Phoenix Section. His current research interests include power system modeling, transient and dynamic stability analysis, critical infrastructure resiliency assessment, and wide area measurements-based protection, monitoring, and control.

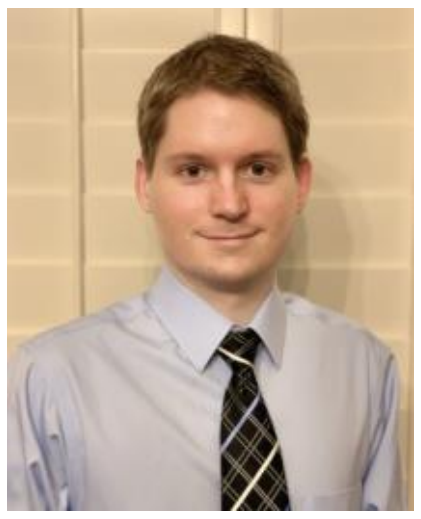

**Trevor Werho** (S’14-M’15) received the degrees of B.S.E., M.S., and Ph.D. degrees in electrical engineering from Arizona State University, Tempe, AZ, USA, in 2011, 2013, and 2015, respectively. He is currently working as a post-doctoral scholar at Arizona State University researching wind and solar forecasting.

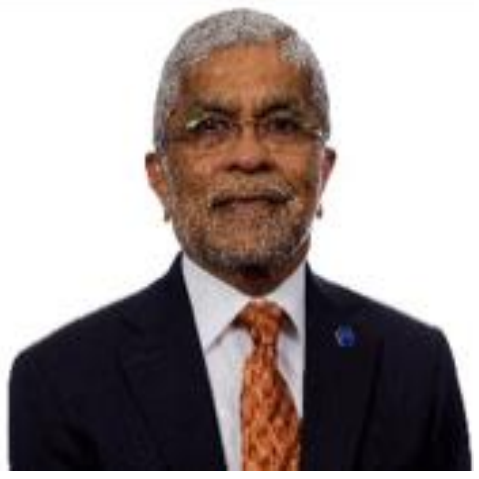

**Vijay Vittal** (S’78-M’82-SM’87–F’97) received the B.E. degree in electrical engineering from the B.M.S. College of Engineering, Bangalore, India, in 1977, the M. Tech. degree from the Indian Institute of Technology, Kanpur, India, in 1979, and the Ph.D. degree from Iowa State University, Ames, in 1982.

He is a Regents’ Professor and the Ira A. Fulton Chair Professor in the Department of Electrical, Computer, and Energy Engineering at Arizona State University, Tempe, AZ. He currently is the Director of the Power System Engineering Research Center (PSERC). Dr. Vittal is a member of the National Academy of Engineering. He is a recipient of the IEEE PES Outstanding Power Engineering Educator award in 2000, the IEEE Herman Halperin T&D Field Award in 2013, and the IEEE PES Prabha S. Kundur Power System Dynamics and Control Award in 2018.